\documentclass[11pt]{article}
\usepackage[centering]{geometry}
\usepackage[utf8]{inputenc}
\usepackage[theoremfont]{newtxtext}
\usepackage{bm}
\usepackage{microtype}
\usepackage{graphicx,xcolor}
\usepackage{comment}
\usepackage{cite}
\usepackage{tikz,tikz-cd}
\usepackage{braket,subcaption,siunitx,slashed,booktabs,enumitem,mathtools,mathdots}
\usepackage{txfonts}
\usepackage[unicode=true,bookmarks=true,bookmarksnumbered=false,bookmarksopen=false,breaklinks=false,pdfborder={0 0 0},pdfborderstyle={},backref=false,colorlinks=true]{hyperref}
\hypersetup{pdftitle={Extended supersymmetry in Dirac action with extra dimensions},pdfauthor={Yukihiro Fujimoto, Kouhei Hasegawa, Kenji Nishiwaki, Makoto Sakamoto, Kentaro Tatsumi, Inori Ueba}}
\DeclareMathOperator{\diag}{diag}%
\newcommand{\one}{\boldsymbol{1}}

\newcommand{\Ncal}{\mathcal{N}}
\newcommand{\ebd}{\boldsymbol{e}}

\begin{document}

\title{Extended supersymmetry in Dirac action with extra dimensions}

\title{
	\vspace{-30pt}
	{\Large \bf Extended supersymmetry in Dirac action with extra dimensions
		\\*[20pt]}
}

\author{
	Yukihiro~Fujimoto,$^a$\footnote{E-mail address: y-fujimoto@oita-ct.ac.jp} \quad
	Kouhei~Hasegawa,$^b$\footnote{E-mail address: kouhei@phys.sci.kobe-u.ac.jp} \quad
	Kenji~Nishiwaki,$^c$\footnote{E-mail address: nishiken@kias.re.kr} \\
	Makoto~Sakamoto,$^b$\footnote{E-mail address: dragon@kobe-u.ac.jp} \quad
	Kentaro~Tatsumi,$^b$\footnote{E-mail address: kentaro@stu.kobe-u.ac.jp} \quad
	Inori~Ueba,$^b$\footnote{E-mail address: i-ueba@stu.kobe-u.ac.jp}\\*[30pt]
	$^a${\it \normalsize National Institute of Technology, Oita College, Oita 870-0152, Japan}\\
	$^b${\it \normalsize Department of Physics, Kobe University, Kobe 657-8501, Japan}\\
	$^c${\it \normalsize School of Physics, Korea Institute for Advanced Study, Seoul 02455, Republic of Korea}\\*[55pt]
}

\date{
	\centerline{\small \bf Abstract}
	\begin{minipage}{0.9\textwidth}
		\medskip\medskip 
		\normalsize
We investigate a new realization of extended quantum-mechanical supersymmetry.
We first show that an $\mathcal{N}=2$ quantum-mechanical supersymmetry is hidden in the four-dimensional (4D) spectrum of the Kaluza--Klein decomposition for the higher dimensional Dirac field, that is, Kaluza--Klein mode functions of 4D right-handed spinors and 4D left-handed ones form $\mathcal{N}=2$ supermultiplets. 
In addition to $\mathcal{N}=2$ supersymmetry, we discover that an $\mathcal{N}$-extended supersymmetry ($\mathcal{N} = d+2\ (d+1)$ for $d=$ even (odd) extra dimensions) is further hidden in the 4D spectrum. The extended symmetry can explain additional degeneracy of the spectrum. 
Furthermore, we show that a superpotential can be introduced into the $\mathcal{N}$-extended supercharges and clarify the condition to preserve the supersymmetry. The partial breaking of the supersymmetry is also demonstrated.
		\begin{flushleft}
		\end{flushleft}
		\begin{flushright}
			\normalsize KIAS-P18035, KOBE-TH-18-01 \\*[55pt]
		\end{flushright}
	\end{minipage}}

	\begin{titlepage}
		\maketitle
		\thispagestyle{empty}
	\end{titlepage}

%
\section{Introduction}
\label{sec:Introduction}

Quantum-mechanical supersymmetry (QM SUSY) has a wide range of applicable topics, e.g. exactly solvable systems in quantum mechanics \cite{Cooper, Quesne2005,Bagchi2009,Odake2009,Fernandez2009}, black holes and AdS/CFT \cite{claus1998,Gibbons1999,pacumio2000,Bellucci2003,Bakas2009}, Sachdev--Ye--Kitaev model \cite{Fu2017,Li2017,Yoon2017,Murugan2017,Jones2018}, 
and so on.
Thus it is interesting and important to investigate new realizations of QM SUSY. 
In particular, it is worth while constructing
$\Ncal$-extended supersymmetry, since not so many models with arbitrary large 
$\Ncal$-extended one are known. 

The $\Ncal$-extended supersymmetry has $\Ncal$ independent supercharges corresponding to the square root of a Hamiltonian, and supercharges relate degenerate eigenstates of the Hamiltonian. 
In order to find a new realization of $\Ncal$-extended QM SUSY, we will investigate fermions in the higher dimensional space-time, because the 4D {spectra} in gauge/gravity theories with extra dimensions have been found to be governed by $\mathcal{N}=2$ QM SUSYs~\cite{Lim2005,Lim2008a,Lim2008b,Nagasawa2011}.
We then expect that $\Ncal=2$ QM SUSY will be hidden also in the 4D spectrum of higher dimensional fermions. 
However, the degeneracy of such 4D spectrum is found to be, in general, much larger than
that expected by the $\Ncal=2$ supersymmetry.
This may imply that there should exist some symmetries in addition to the
$\Ncal=2$ supersymmetry.
The main purpose of this letter is to show that an $\Ncal$-extended QM SUSY
is hidden in the 4D spectrum of the higher dimensional Dirac action
and can explain the degeneracy of the 4D spectrum.

In this paper, we show that there generally exists an $\mathcal{N}=2$ QM SUSY hidden in the 4D spectrum of a higher dimensional Dirac action after the compactification of extra space dimensions.
The symmetry turns out to connect the 4D right-handed spinors and the 4D left-handed ones. Furthermore, we find an {$\mathcal{N}$}-extended QM SUSY with $\mathcal{N} = d+2\ (d+1)$ for $d=$ even (odd) extra dimensions.
We then show that the $\mathcal{N}$-extended supersymmetry explains the degeneracy of the 4D spectrum in the higher dimensional Dirac action.
Explicit constructions of $\mathcal{N}$-extended supersymmetry algebras with
higher $\mathcal{N}$, have been investigated~\cite{Toppan2001,Kuznetsova2006,Faux2005,DeCrombrugghe1983,Howe1988,Akulov1999,Nagasawa2004,Nagasawa2005}, but what we found in this
paper gives a new realization of the $\mathcal{N}$-extended supersymmetry
algebra.

This paper is organized as follows:
In Section~\ref{sec:Action-and-KK}, we clarify the properties of the mode functions of the Dirac field providing the 4D mass eigenstates. In Section~\ref{sec:N2-SQM}, we reveal a hidden $\Ncal=2$ QM SUSY on such mode functions. We show a hidden $\mathcal{N}$-extended QM SUSY in Section~\ref{sec:extended-SQM}. 
It is shown that the whole degenerate mode functions on the hyperrectangular internal space (extra dimensions) can be explained by the $\mathcal{N}$-extended supersymmetry.
In section \ref{sec:Extension-superpotential}, we introduce a superpotential into the extended QM SUSY. 
Then, we clarify the conditions that the superpotential can preserve or (partially)
break the extended supersymmetry.
Section~\ref{sec:summary} is devoted to summary and discussion.

%
\section{Higher dimensional Dirac action and KK decomposition}
\label{sec:Action-and-KK}

In this section, we discuss the 
Kaluza-Klein (KK) decomposition of a higher dimensional Dirac field and clarify the properties of the mode functions of the KK
decomposition. The decomposition is executed in such a way that
the induced Dirac action at {four dimensions (4D)} is constructed
in terms of the 4D mass eigenstates.

Let us consider the action of a free Dirac field on the direct product of the 4D Minkowski space-time $M^{4}$ and a $d$-dimensional flat internal space $\Omega$ given as\footnote{For earlier works on higher dimensional spinors, see e.g. \cite{Brauer1935,Wetterich1982,Kugo:1982bn}.}
\begin{align}
S & =\int_{M^{4}}d^{4}x\int_{\Omega}d^{d}y\,\bar{\Psi}(x,y)(i\Gamma^{\mu}\partial_{\mu}+i\Gamma^{y_{k}}\partial_{y_{k}}-M)\Psi(x,y)\,.\label{eq:action}
\end{align}
The $x^{\mu}$ ($\mu=0,1,2,3$) are the coordinates of the Minkowski space-time $M^{4}$ and $y_{k}$ ($k=1, 2, \cdots, d$) are the coordinates of the internal space $\Omega$. We take the metric as $\eta_{NM}=\eta^{NM}=\diag(-1,+1,{\cdots},+1)$ where $N,M=0,1,{\cdots},y_{d}$.
The gamma matrices $\Gamma^N$ satisfy {the} Clifford algebra,
\begin{align}
	\{\Gamma^N,\Gamma^M\}&=-2\eta^{NM}\one_{2^{\lfloor d/2\rfloor}}\,.
	\label{eq:Clifford_Definition}
\end{align}
The $\Psi(x,y)$ is the $(d+4)$-dimensional Dirac spinor which has $2^{\lfloor d/2+2\rfloor}$ components with mass $M$. The symbol $\lfloor d/2+2\rfloor$ is the maximum integer less than or equal to $d/2+2$. 
$\one_{n}$ represents the $n\times n$ unit matrix.
The Dirac conjugate is defined as $\bar{\Psi}(x,y)=\Psi^{\dagger}(x,y)\Gamma^{0}$.
It turns out to be very convenient for
our analysis to use the following representation of the gamma matrices
$\Gamma^{N}$, based on the 4D chiral representation:
%
\begin{align}
\Gamma^{0} & =\sigma^{1}\otimes\one_{2^{\lfloor d/2\rfloor}}\otimes\one_{2}\,,  \notag\\
\Gamma^{a} & =i\sigma^{2}\otimes\one_{2^{\lfloor d/2\rfloor}}\otimes\sigma^{a} &&(a=1,2,3)\,,   \notag\\
\Gamma^{y_{k}} & =-\sigma^{3}\otimes\gamma^{y_{k}}\otimes\one_{2} && (k=1,2,\cdots,d)\,,
  \label{eq:gamma_rep}
\end{align}
%
where $\sigma^{a}$ denote the Pauli matrices and $\gamma^{y_{k}}$ ($k=1,2,\cdots,d$)
correspond to the internal space gamma matrices which satisfy $\{\gamma^{y_{k}},\gamma^{y_{h}}\}=-2\delta^{kh}\one_{2^{\lfloor d/2\rfloor}}$
with $(\gamma^{y_{k}})^{\dagger}=-\gamma^{y_{k}}$ (see Eq.~(\ref{eq:Clifford_Definition})).\footnote{For the case of $d=1$, $\gamma^{y_1}$ is defined as $i$.} We define the 4D chiral matrix $\Gamma_{5}$ and {the internal space chiral matrix} $\Gamma_{d+1}$ (for even $d$) by\footnote{In odd dimensions, there is no internal chiral matrix corresponding to $\Gamma_{d+1}$ or $\gamma_{d+1}$.}
%
\begin{alignat}{3}
\Gamma_{5} & =i\Gamma^{0}\dots\Gamma^{3} && =-\sigma^{3}\otimes\one_{2^{\lfloor d/2\rfloor}}\otimes\one_{2}\,,  \notag\\
\Gamma_{d+1} & =i^{d/2}\Gamma^{y_{1}}\dots\Gamma^{y_{d}}&&=\one_{2}\otimes\gamma_{d+1}\otimes\one_{2} &&\;(\text{for even }d)\,,
\end{alignat}
%
with $\gamma_{d+1}=i^{d/2}\gamma^{y_{1}}\dots\gamma^{y_{d}}$. The
4D chiral projection matrices $P_{R/L}$ are given by
%
\begin{align}
P_{L}=\frac{\one-\Gamma_{5}}{2} & =\begin{pmatrix}
1 & 0\\
0 & 0
\end{pmatrix}\otimes\one_{2^{\lfloor d/2\rfloor}}\otimes\one_{2}\,,  \notag\\
P_{R}=\frac{\one+\Gamma_{5}}{2} & =\begin{pmatrix}
0 & 0\\
0 & 1
\end{pmatrix}\otimes\one_{2^{\lfloor d/2\rfloor}}\otimes\one_{2}\,.
\end{align}
%
The eigenstates of the 4D chiral matrix $\Gamma_{5}$ with the eigenvalue
$+1$ ($-1$) are called right-handed (left-handed).

In terms of the 4D left-handed (right-handed) two component chiral
spinors $\phi_{\alpha}^{(n)}(x)$ ($\bar{\chi}_{\alpha}^{(n)}(x)$),\footnote{For two component spinors, we follow the notation adopted in~\cite{WessBagger}, which includes the followings:
${\sigma}^{\mu} = \{ \one_2, \sigma^a \}$,
$\bar{\sigma}^{\mu} = \{ \one_2, -\sigma^a \}$,
$(\phi_{\alpha}^{(n)}(x))^\dagger = \bar{\phi}_{\alpha}^{(n)}(x)$, and
$(\bar{\chi}_{\alpha}^{(n)}(x))^\dagger = {\chi}_{\alpha}^{(n)}(x)$.
}
the KK decomposition of the ($d+4$)-dimensional Dirac field $\Psi(x,y)$
will be given by 
%
\begin{align}
\Psi(x,y)
  & =\sum_{n}\sum_{\alpha}
   \Big\{\,\ebd_{L}\otimes\boldsymbol{f}_{\alpha}^{(n)}(y)\otimes\phi_{\alpha}^{(n)}(x)
         +\ebd_{R}\otimes\boldsymbol{g}_{\alpha}^{(n)}(y) 
         \otimes\bar{\chi}_{\alpha}^{(n)}(x) \,\Big\} \notag\\
 & =\sum_{n}\sum_{\alpha}\begin{pmatrix}\boldsymbol{f}_{\alpha}^{(n)}(y)\otimes\phi_{\alpha}^{(n)}(x) \\
\boldsymbol{g}_{\alpha}^{(n)}(y)\otimes\bar{\chi}_{\alpha}^{(n)}(x)
\end{pmatrix}\,,
 \label{eq:expansion-Dirac}
\end{align}
%
where $\ebd_{L}=(1,0)^{\text{T}}$ and $\ebd_{R}=(0,1)^{\text{T}}$
characterize the 4D chirality. The index $n$ represents the $n$-th
level of the KK modes and $\alpha$ denotes the index that distinguishes
the degeneracy of the $n$-th KK modes (if exist). The mode functions
$\boldsymbol{f}_{\alpha}^{(n)}(y)$ ($\boldsymbol{g}_{\alpha}^{(n)}(y)$)
have $2^{\lfloor d/2\rfloor}$ components and are assumed to form
a complete set with respect to the internal space associated with
the 4D left-handed (right-handed) chiral spinors $\phi_{\alpha}^{(n)}(x)$
($\bar{\chi}_{\alpha}^{(n)}(x)$).

Substituting this expansion~\eqref{eq:expansion-Dirac} into the action~\eqref{eq:action}, we find
%
\begin{align}
S =\int_{M_{4}}d^{4}x\,\sum_{\mathclap{n,m}}\sum_{\mathclap{\alpha,\beta}}
\bigg[&\braket{\boldsymbol{f}_{\alpha}^{(n)}|\boldsymbol{f}_{\beta}^{(m)}}\bar{\phi}_{\alpha}^{(n)}(x)i\bar{\sigma}^{\mu}\partial_{\mu}\phi_{\beta}^{(m)}(x)
 +\braket{\boldsymbol{g}_{\alpha}^{(n)}|\boldsymbol{g}_{\beta}^{(m)}}\chi_{\alpha}^{(n)}(x)i{\sigma}^{\mu}\partial_{\mu}\bar{\chi}_{\beta}^{(m)}(x)\nonumber \\
 & +\braket{\boldsymbol{f}_{\alpha}^{(n)}|A\boldsymbol{g}_{\beta}^{(m)}}\bar{\phi}_{\alpha}^{(n)}(x)\bar{\chi}_{\beta}^{(m)}(x)
  +\braket{\boldsymbol{g}_{\alpha}^{(n)}|A^{\dagger}\boldsymbol{f}_{\beta}^{(m)}}\chi_{\alpha}^{(n)}(x)\phi_{\beta}^{(m)}(x)\bigg]\,,\label{eq:decomposed-action}
\end{align}
%
where $\braket{\boldsymbol{X}|\boldsymbol{Y}}=\int_{\Omega}d^{d}y\,\boldsymbol{X}^{\dagger}(y)\boldsymbol{Y}(y)$
and $A=+i\gamma^{{y_k}}\partial_{{y_k}}-M$, $A^{\dagger}=-i\gamma^{{y_k}}\partial_{{y_k}}-M$.
We note that $A$ and $A^{\dagger}$ are the off-diagonal components
of the Dirac operator $\Gamma^{0}(i\Gamma^{y_{k}}\partial_{y_{k}}-M)$
written as
%
\begin{align}
\Gamma^{0}(i\Gamma^{y_{k}}\partial_{y_{k}}-M) & =\begin{pmatrix}0 & A\\
A^{\dagger} & 0
\end{pmatrix}\otimes\one_{2}\,.
\end{align}
%
%
We require that $\phi_{\alpha}^{(n)}(x)$ and $\bar{\chi}_{\alpha}^{(n)}(x)$
are mass eigenstates of the 4D spectrum and that the action~\eqref{eq:decomposed-action}
should be written into the form
%
\begin{align}
S=\int_{M^{4}}dx\,\bigg\{\,&\sum_{\alpha}\sum_{n}\bar{\psi}_{\alpha}^{(n)}(x)(i\gamma^{\mu}\partial_{\mu}+m_{n})\psi_{\alpha}^{(n)}(x)\nonumber\\
&+\sum_{\alpha}\bar{\phi}_{\alpha}^{(0)}(x)i\bar{\sigma}^{\mu}\partial_\mu\phi_{\alpha}^{(0)}(x)+\chi_{\alpha}^{(0)}(x)i{\sigma}^{\mu}\partial_\mu\bar{\chi}_{\alpha}^{(0)}(x)\,\bigg\}\,,
\end{align}
%
where $\psi_{\alpha}^{(n)}(x)=\big(\phi_{\alpha}^{(n)}(x),\bar{\chi}_{\alpha}^{(n)}(x)\big)^{\text{T}}$
are the $n$-th KK 4D Dirac fields with mass $m_{n}$ and $\phi_{\alpha}^{(0)}(x)$
($\bar{\chi}_{\alpha}^{(0)}(x)$) in the second line are massless
left-handed (right-handed) 4D chiral fields. This can be realized,
provided the mode functions $\boldsymbol{f}_{\alpha}^{(n)}(y)$
and $\boldsymbol{g}_{\alpha}^{(n)}(y)$ satisfy the following orthonormality
relations:
%
\begin{align}
&
\braket{\boldsymbol{f}_{\alpha}^{(n)}|\boldsymbol{f}_{\beta}^{(m)}}=\braket{\boldsymbol{g}_{\alpha}^{(n)}|\boldsymbol{g}_{\beta}^{(m)}}=\delta_{\alpha\beta}\delta^{nm}\,, \notag\\
&\hspace{-5mm}
\braket{\boldsymbol{f}_{\alpha}^{(n)}|A\boldsymbol{g}_{\beta}^{(m)}}=\braket{\boldsymbol{g}_{\alpha}^{(n)}|A^{\dagger}\boldsymbol{f}_{\beta}^{(m)}}=m_{n}\delta_{\alpha\beta}\delta^{nm}\,.
\label{eq:ortho}
\end{align}
%

%
\section{$\Ncal=2$ quantum-mechanical supersymmetry}
\label{sec:N2-SQM}

In this section, we show that the orthonormality relations~\eqref{eq:ortho} imply that the mode functions $\boldsymbol{f}_{\alpha}^{(n)}(y)$ and $\boldsymbol{g}_{\alpha}^{(n)}(y)$ form supersymmetric partners (except for zero modes) in an $\Ncal=2$ {supersymmetric quantum mechanics~(SQM)}\@.

Since the mode functions $\boldsymbol{f}_{\alpha}^{(n)}(y)$ and $\boldsymbol{g}_{\alpha}^{(n)}(y)$ are assumed to form complete sets on the internal space, Eq.~\eqref{eq:ortho} {leads to} the relations
%
\begin{align}
A^{\dagger}\boldsymbol{f}_{\alpha}^{(n)}(y) & =m_{n}\boldsymbol{g}_{\alpha}^{(n)}(y)\,, & A\boldsymbol{g}_{\alpha}^{(n)}(y) & =m_{n}\boldsymbol{f}_{\alpha}^{(n)}(y)\,.\label{eq:SQM_relation_in_conponents}
\end{align}
%
It immediately follows that $\boldsymbol{f}_{\alpha}^{(n)}(y)$ and $\boldsymbol{g}_{\alpha}^{(n)}(y)$ satisfy the eigenvalue equations
%
\begin{align}
(-\partial_{y}^{2}+M^{2})\boldsymbol{f}_{\alpha}^{(n)}(y) & =m_{n}^{2}\boldsymbol{f}_{\alpha}^{(n)}(y)\,,  \notag\\
(-\partial_{y}^{2}+M^{2})\boldsymbol{g}_{\alpha}^{(n)}(y) & =m_{n}^{2}\boldsymbol{g}_{\alpha}^{(n)}(y)\,,
\label{eq:eigen_eq}
\end{align}
%
and that the above relations~\eqref{eq:SQM_relation_in_conponents} and~\eqref{eq:eigen_eq}
can be naturally embedded in a system of an $\Ncal=2$ SQM, as explained below. 

Let us introduce the supercharge $Q$ and the ``fermion'' number operator $(-1)^{F}$ such as
%
\begin{align}
Q & =\begin{pmatrix}0 & A\\
A^{\dagger} & 0
\end{pmatrix}\,, & (-1)^{F} & =\begin{pmatrix}-\one_{2^{\lfloor d/2\rfloor}} & 0\\
0 & \one_{2^{\lfloor d/2\rfloor}}
\end{pmatrix}\,,
\end{align}
and then define the Hamiltonian by
\begin{align}
H & =Q^{2}\,.
\end{align}
%
This system is known as the $\Ncal=2$ {SQM} (see reviews~\cite{Cooper}).\footnote{If one defines $Q_{1}=Q$ and $Q_{2}=i(-1)^{F}Q$, then they form the $\Ncal=2$ supersymmetry algebra $\{Q_{j},Q_{k}\}=2H\delta_{jk}$
for $j,k=1,2$.}
It should be noticed that the supercharge $Q$ and the fermion number
operator $(-1)^{F}$ are represented as
%
\begin{align}
\Gamma^{0}(i\Gamma^{y}\partial_{y}-M) & =Q\otimes\one_{2}\,,&
\Gamma_{5} & =(-1)^{F}\otimes\one_{2}\,.\label{eq:relation-Q-and-action}
\end{align}
%
It follows that $Q$ and $(-1)^{F}$ can be interpreted as the Dirac operator on the internal space and the 4D chiral operator, respectively.

To rewrite the relations~\eqref{eq:SQM_relation_in_conponents} and~\eqref{eq:eigen_eq} in the language of the $\Ncal=2$ SQM, we introduce
%
\begin{align}
\Phi_{\alpha, +}^{(n)}(y) 
 & =\begin{pmatrix}
           0 \\
           \boldsymbol{g}_{\alpha}^{(n)}(y)
    \end{pmatrix}\,,
 & \Phi_{\alpha, -}^{(n)}(y)
 & =\begin{pmatrix}\boldsymbol{f}_{\alpha}^{(n)}(y)\\
                     0
    \end{pmatrix}\,.
\label{eq:SQM_relation-1}
\end{align}
%
Then, in terms of $\Phi_{\alpha, \pm}^{(n)}(y)$, the relations~\eqref{eq:SQM_relation_in_conponents} and~\eqref{eq:eigen_eq} can be rewritten as
%
\begin{align}
Q\Phi_{\alpha, \pm}^{(n)}(y) & =m_{n}\Phi_{\alpha, \mp}^{(n)}(y)\,, \label{eq:SQM_rel}\\
H\Phi_{\alpha, \pm}^{(n)}(y) & =m_{n}^{2}\Phi_{\alpha, \pm}^{(n)}(y)\,, \\
\intertext{with}(-1)^{F}\Phi_{\alpha, \pm}^{(n)}(y) & =\pm\Phi_{\alpha, \pm}^{(n)}(y)\,.
\end{align}
%

\begin{figure}[th]\centering%
	\includegraphics{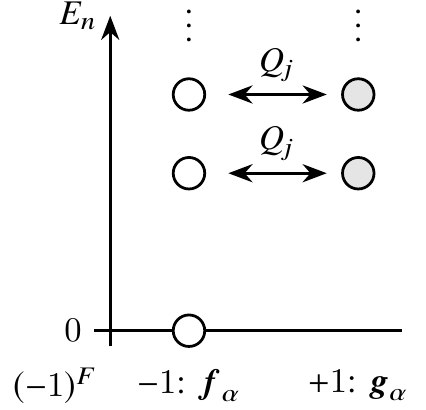}%
	\caption{The one dimensional SQM model (known as a Witten model~\cite{Witten1981}) 
	has the doubly degenerated spectrum except for the zero energy state ($j=1,2$).}
	\label{fig:spec-1}
\end{figure}

%
We call Eqs.~\eqref{eq:SQM_relation_in_conponents} and~\eqref{eq:SQM_rel},
supersymmetric relations. The mode functions 
$\Phi_{\alpha, +}^{(n)}(y)$ (or $\boldsymbol{g}_{\alpha}^{(n)}(y)$)
and $\Phi_{\alpha, -}^{(n)}(y)$ (or $\boldsymbol{f}_{\alpha}^{(n)}(y$)) are
 (at least) doubly degenerate in the spectrum except for
zero energy states (see Figure~\ref{fig:spec-1}), and they form
$\Ncal=2$ supermultiplets.
Thus, the $\mathcal{N}=2$ supersymmetry turns out to be hidden
in the 4D spectrum of the higher dimensional Dirac theory, 
as well as higher dimensional gauge/gravity theories~\cite{Lim2005,Lim2008a,Lim2008b,Nagasawa2011}.

\begin{figure}[th]\centering%
	\includegraphics{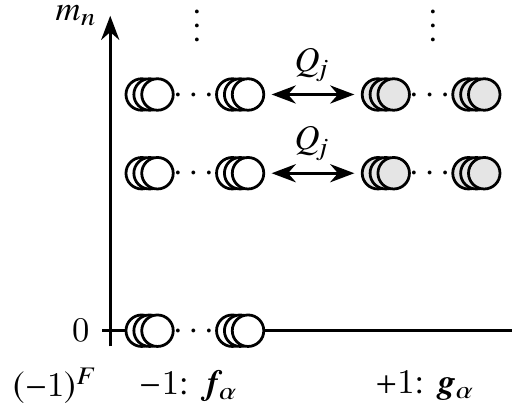}%
	\caption{The eigenfunctions 
	$\boldsymbol{f}_{\alpha}^{(n)}(y)$ and $\boldsymbol{g}_{\alpha}^{(n)}(y)$
	of the higher dimensional Dirac field are, in general, multiply
	degenerate at each KK level.
	An $\mathcal{N}$-extended supersymmetry {with the supercharges $Q_j$
	($j=1,2,\cdots,d+2$ ($j=1,2,\cdots,d+1$) for even (odd) dimensions)}
	is hidden in the spectrum.}
	\label{fig:spec-2}
\end{figure}

Here, a question arises about the degeneracy of the spectrum. Although
the $\Ncal=2$ supersymmetry assures a pair of eigenstates in the
spectrum, as seen in Figure~\ref{fig:spec-1}, the degeneracy of
the spectrum in the higher dimensional Dirac action is found 
to be much larger than that expected by the
$\Ncal=2$ supersymmetry (see Figure~\ref{fig:spec-2}), in general.
This fact suggests that there should exist some symmetries in addition
to the $\Ncal=2$ supersymmetry. To reveal hidden symmetries on the degeneracy of the spectrum is the purpose of the next
section.

{Before closing this section, we should give a comment on the Hermiticity
property of the supercharge. The supercharge $Q$ has to be Hermitian
in the $\Ncal=2$ SQM but the Hermiticity of $Q$ would not be trivial
if the internal space $\Omega$ has boundaries. To verify the Hermiticity
of $Q$, let us examine the variational principle of the action, $\delta S=0$,
which leads to the ($d+4$)-dimensional Dirac equation
and also the condition for the surface integral, i.e.
%
\begin{align}
\int_{\partial\Omega}d^{d-1}y\,\begin{pmatrix}\boldsymbol{f}_{\tilde{\alpha}}^{(\tilde{n})}(y)\\
\boldsymbol{g}_{\tilde{\beta}}^{({\tilde{m}})}(y)
\end{pmatrix}^{\dagger}\begin{pmatrix}0 & in_{y_k}\gamma^{y_k}\\
-in_{y_k}\gamma^{y_k} & 0
\end{pmatrix}\begin{pmatrix}\boldsymbol{f}_{\alpha}^{(n)}(y)\\
\boldsymbol{g}_{\beta}^{({m})}(y)
\end{pmatrix} & =0{\,,} \label{eq:non-local_BC}
\end{align}
%
for any $n,m,\alpha,\beta$ and $\tilde{n},\tilde{m},\tilde{\alpha},\tilde{\beta}$.
The $n_{y_k}$ is a normal vector at each point of the boundary $\partial\Omega$.
It turns out that the supercharge $Q$ is Hermitian, provided that 
boundary conditions on 
$\boldsymbol{f}_{\alpha}^{(n)}(y)$ and $\boldsymbol{g}_{\alpha}^{(n)}(y)$
are chosen to satisfy Eq.~\eqref{eq:non-local_BC}.}
%
%
%

%
\section{ $\mathcal{N}$-extended quantum-mechanical supersymmetry}
\label{sec:extended-SQM}

{Interestingly,}
{the previous $\Ncal = 2$ supersymmetry can be enlarged to an $\Ncal$-extended supersymmetry describing}
the additional
degeneracy of the spectrum. The supersymmetry has $d+2$ ($d+1$)
supercharges for $d=$ even (odd) extra dimensions. 
The supercharges form the $\Ncal$-extended supersymmetry algebra such as
%
\begin{align}
\{Q_{i},Q_{j}\} & =2H\delta_{ij}\,, & [Q_{i},H] & =0\,,\label{eq:extended-SQM}
\end{align}
%
where $i,j=1,2,\cdots,d+2$ ($i,j=1,2,\cdots,d+1$) for even (odd) dimensions.
The Hamiltonian $H$ is the same as that given in the $\Ncal=2$ supersymmetry algebra. 
The supercharges
are defined by the composition of the reflections operators as
%
\begin{align}
Q_{k} &=i(-1)^{F}(\one_{2}\otimes\gamma_{d+1}\gamma^{y_{k}})R_{k}Q
         && (k=1,\cdots,d)\,, \notag\\
Q_{d+1} &=i(-1)^{F}(\one_{2}\otimes\gamma_{d+1})PQ\,, \notag\\
Q_{d+2} &=Q\,,
\label{N-extended-even}
%
\intertext{for even dimensions, and}
%
Q_{k} &=i(-1)^{F}(\one_{2}\otimes i\gamma^{y_{d}}\gamma^{y_{k}})R_{d}R_{k}Q
         &&  (k=1,\cdots,d-1)\,,  \notag\\
Q_{d} &=i(-1)^{F}(\one_{2}\otimes i\gamma^{y_{d}})R_{d}PQ\,,   \notag\\
Q_{d+1} &=Q\,,
\label{N-extended-odd}
\end{align}
%
for odd dimensions.
The $R_{k}$ ($k=1,2,\cdots,d$) denotes  
the reflection operator for the $y_{k}$ direction, i.e. 
$(R_{k}f)(y_{1},\cdots,y_{k},\cdots,y_{d}) \equiv f(y_{1},\cdots,-y_{k},\cdots,y_{d})$
for any function $f(y_{1},\cdots,y_{d})$.\footnote{
We note that the reflection operator $R_{k}$ acts on $\partial_{l}$ as
$R_k \partial_{y_l} R_k^{-1} = (1-2\delta_{kl})\partial_{y_l}$.
}
$P=\prod_{k=1}^{d}R_{k}$ represents the point reflection (or parity) operator
of the internal space. 
Here (when $d$ is more than one), the element $i (-1)^F Q$ is not included in this algebra since it is commutative with the other components except for $Q$.

%

%
To demonstrate that the supercharges $Q_{j}$ can explain the degeneracy
of the spectrum at each KK level with $m^{2}_{n} > 0$,
let us consider, as an example, the $d$-dimensional hyperrectangle 
internal space:
%
%
\begin{align}
\Omega & =[-L_{1}/2, L_{1}/2]\times\dots\times[-L_{d}/2,L_{d}/2]\,,
\end{align}
%
where $L_{k}$ ($k=1,2,\cdots,d$) is the length of the $k$-th side of the hyperrectangle with
the Dirichlet boundary condition on $\boldsymbol{g}_{\alpha}^{(n)}(y)$, i.e.
%
\begin{align}
\boldsymbol{g}_{\alpha}^{(n)}(y) & =0 &\big( \text{or}\ \ 
\Phi_{\alpha, +}^{(n)}(y) &= 0\,\big) & \text{on}\ \partial\Omega\,.
\label{eq:BC_g}
\intertext{Then, the supersymmetric relations~\eqref{eq:SQM_relation_in_conponents} (or~\eqref{eq:SQM_rel})
lead to the following boundary condition on $\boldsymbol{f}_{\alpha}^{(n)}(y)$~\cite{Fujimoto2017a,Fujimoto2017b}:}
A^{\dagger}\boldsymbol{f}_{\alpha}^{(n)}(y) & =0
&\big( \text{or}\ \ Q\Phi_{\alpha,-}^{(n)}(y)&= 0\,\big) & \text{on}\ \partial\Omega\,.
\label{eq:BC_f}
\end{align}
%
It should be emphasized that the boundary conditions~\eqref{eq:BC_g} and~\eqref{eq:BC_f} are compatible with the
$\mathcal{N}$-extended supersymmetry.

The $n$-th mode functions $\boldsymbol{g}_{\alpha}^{(n)}(y)$
and $\boldsymbol{f}_{\alpha}^{(n)}(y)$ with a KK mass $m_{n}^{2}>0$
are found to be of the form
%
\begin{align}
\boldsymbol{g}_{\alpha}^{(n)}(y) & =h^{(n)}(y)\boldsymbol{e}_{\alpha}\,, &  \\
\boldsymbol{f}_{\alpha}^{(n)}(y) & =\frac{1}{m_{n}}A\boldsymbol{g}_{\alpha}^{(n)}(y)
& \left( \text{or} \ \ \Phi_{\alpha,-}^{(n)}(y) =\frac{1}{m_{n}} Q \Phi_{\alpha,+}^{(n)}(y) \right) \,,
	\label{eq:SUSY_relation_with_Q}
\end{align}
%
where $h^{(n)}(y)$ is a scalar function.
%
{Hereafter, we adopt the following variables for representing $\alpha$ concretely, where the} basis vectors of the spinor space 
$\boldsymbol{e}_{\alpha} = \boldsymbol{e}_{s_{1} s_{2} \cdots s_{{\lfloor d/2\rfloor}}}$
($s_{p}=\pm$ for $p=1,2,\cdots,\lfloor d/2\rfloor$) are the eigenvectors
of $\gamma_{(p)}=i\gamma^{y_{2p-1}}\gamma^{y_{2p}}$~\cite{Polchinski1-String,Polchinski2-String}, i.e.
%
\begin{align}
\gamma_{(p)}\boldsymbol{e}_{s_{1} \cdots s_{p} \cdots s_{{\lfloor d/2\rfloor}}}
& = s_{p}\boldsymbol{e}_{s_{1} \cdots s_{p} \cdots s_{{\lfloor d/2\rfloor}}} \,,
\end{align}
%
for all $p=1,2,\cdots,\lfloor d/2\rfloor$.\footnote{
{Note that $\boldsymbol{g}_{s_{1} s_{2} \cdots s_{{\lfloor d/2\rfloor}}}^{(n)}$ are eigenvectors of $\gamma_{(p)}$, while $\boldsymbol{f}_{s_{1} s_{2} \cdots s_{{\lfloor d/2\rfloor}}}^{(n)}$ are not.
The forms of $\boldsymbol{f}_{s_{1} s_{2} \cdots s_{{\lfloor d/2\rfloor}}}^{(n)}$ are determined by the corresponding states of $\boldsymbol{g}_{s_{1} s_{2} \cdots s_{{\lfloor d/2\rfloor}}}^{(n)}$ through the relation~(\ref{eq:SUSY_relation_with_Q}).}
}
{$s_p$ represents an eigenvalue of the $p$-th internal chirality {of $\gamma_{(p)}$}.}
Thus, the degeneracy at each KK level is $2 \times 2^{\lfloor d/2\rfloor}$.
(The additional factor $2$ corresponds to the degeneracy between
$\boldsymbol{g}_{\alpha}^{(n)}$ and $\boldsymbol{f}_{\alpha}^{(n)}$.)
The scalar function $h^{(n)}(y)$
is the eigenmode of the Hamiltonian $H=-\partial_{y}^{2}+M^{2}$
with the eigenvalue $m_{n}^{2}=M^{2}+\sum_{k=1}^{d}(n_{k}\pi)^{2}/L_{k}^{2}$
where $n_{k}=1,2,\cdots$ ($k=1,2,\cdots,d$),
and is explicitly given by
%
\begin{align}
h^{(n)}(y) & 
 =\prod_{k=1}^{d}
   \sqrt{\frac{2}{L_{k}}}\sin\left(\frac{n_{k}\pi}{L_{k}}
     \Big(y_{k}+\frac{L_{k}}{2}\Big)\right)\,.
     \label{eq:h-function}
\end{align}
%

\begin{figure}[th]\centering%
	\includegraphics{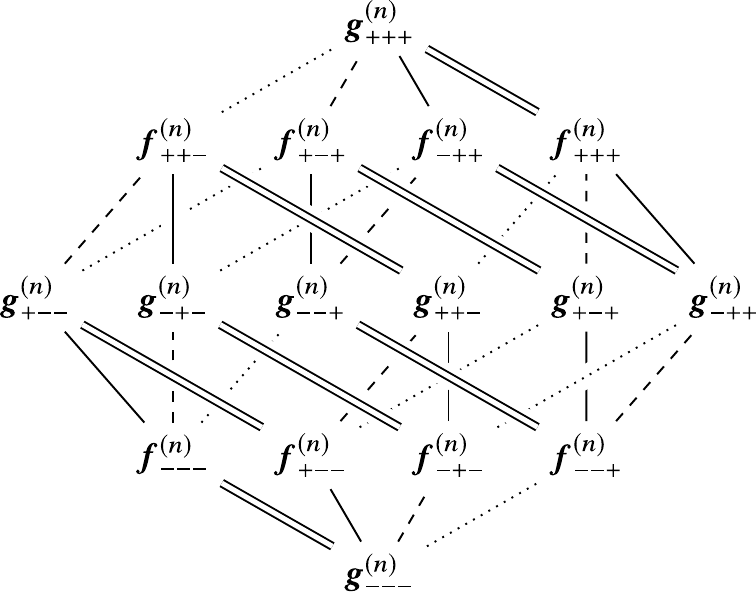}%
\caption{
The $\boldsymbol{g}_{s_{1}s_{2}s_{3}}^{(n)}$ and $\boldsymbol{f}_{s_{1}s_{2}s_{3}}^{(n)}$
$(s_{1},s_{2}.s_{3}=\pm)$ are the degenerate $n$-th mode functions 
on the six-dimensional hyperrectangle in the ten-dimensional space-time.
The $\Ncal$-extended supercharges generate the whole mode functions. 
The lines indicate the relationships through the $\Ncal$-extended supercharges,
where the double continuous, single continuous, dashed, and dotted lines correspond to
$Q_7$ or $Q_8$,
$Q_1$ or $Q_2$,
$Q_3$ or $Q_4$, and
$Q_5$ or $Q_6$, respectively.
}
\label{fig:10d}
\end{figure}

%
It is easy to verify that the $\Ncal$-extended supercharges $Q_{j}$ 
can explain the degeneracy of the mode functions, as we expected.
{We can derive the following general relationships with $k' = 2p-1 \text{ or } 2p$ {($p=1,2,\cdots,\lfloor d/2\rfloor$)} for $d$ being even,
\begin{align}
Q_{k'} \, \Phi_{s_{1} \cdots s_{p} \cdots s_{{\lfloor d/2\rfloor}}, \pm}^{(n)}(y) &\propto
\Phi_{s_{1} \cdots (-s_{p}) \cdots s_{{\lfloor d/2\rfloor}}, \mp}^{(n)}(y)\,, 
\label{eq:Q_transform_1} \\
Q_{d+1,d+2} \, \Phi_{s_{1} \cdots s_{p} \cdots s_{{\lfloor d/2\rfloor}}, \pm}^{(n)}(y) &\propto
\Phi_{s_{1} \cdots s_{p} \cdots s_{{\lfloor d/2\rfloor}}, \mp}^{(n)}(y)\,,
\end{align}
and for $d$ being odd,
\begin{align}
Q_{k'} \, \Phi_{s_{1} \cdots s_{p} \cdots s_{{\lfloor d/2\rfloor}}, \pm}^{(n)}(y) &\propto
\Phi_{s_{1} \cdots (-s_{p}) \cdots s_{{\lfloor d/2\rfloor}}, \mp}^{(n)}(y)\,, 
\label{eq:Q_transform_2} \\
Q_{d,d+1} \, \Phi_{s_{1} \cdots s_{p} \cdots s_{{\lfloor d/2\rfloor}}, \pm}^{(n)}(y) &\propto
\Phi_{s_{1} \cdots s_{p} \cdots s_{{\lfloor d/2\rfloor}}, \mp}^{(n)}(y)\,,
\end{align}
where flips are observed in one of the internal chiralities {for the cases of (\ref{eq:Q_transform_1}) and (\ref{eq:Q_transform_2})}.
Here, we adopt the notation (being similar to that in Eq.~(\ref{eq:SQM_relation-1}))
\begin{align}
\Phi_{s_{1} \cdots s_{p} \cdots s_{{\lfloor d/2\rfloor}}, +}^{(n)}(y) 
 & =\begin{pmatrix}
           0 \\
           \boldsymbol{g}_{s_{1} \cdots s_{p} \cdots s_{{\lfloor d/2\rfloor}}}^{(n)}(y)
    \end{pmatrix}\,, \notag \\
\Phi_{s_{1} \cdots s_{p} \cdots s_{{\lfloor d/2\rfloor}}, -}^{(n)}(y)
 & =\begin{pmatrix}\boldsymbol{f}_{s_{1} \cdots s_{p} \cdots s_{{\lfloor d/2\rfloor}}}^{(n)}(y)\\
                     0
    \end{pmatrix}\,.
\end{align}
We summarize the above statements as follows.
Once} a single mode function 
$\boldsymbol{g}_{\alpha}^{(n)}(y)$ (or $\boldsymbol{f}_{\alpha}^{(n)}(y)$)
is given (for fixed $\alpha$), the whole mode functions can be obtained
by successively operating the $\Ncal$-extended supercharges 
on $\boldsymbol{g}_{\alpha}^{(n)}(y)$ 
(or $\boldsymbol{f}_{\alpha}^{(n)}(y)$) as shown at Figure \ref{fig:10d}.

\section{Extension of supersymmetry with superpotential}
\label{sec:Extension-superpotential}
%

Let us consider the extension of the supersymmetry with a superpotential.
To this end, we replace the bulk mass $M$ with a scalar function $W(y)$.
Then, we can show that the $\mathcal{N}=2$ supersymmetry given in Section 3
holds with the supercharge
%
\begin{align}
Q = \begin{pmatrix}
      0 & i\gamma^{y_{k}}\partial_{y_{k}} - W(y) \\
      -i\gamma^{y_{k}}\partial_{y_{k}} - W(y) & 0
    \end{pmatrix}\,.
\end{align}
%
It is, however, non-trivial for the whole of the extended supersymmetry
to preserve with the introduction of the superpotential $W(y)$.

For $d=$ even, the conditions for a realization of the full $\mathcal{N}=d+2$
extended supersymmetry are found to be given by
%
\begin{align}
\big(R_{k}W\big)(y) = W(y) \qquad \textrm{for}\ k=1,\cdots,d\,,
\label{W(-y)=W(y)}
\end{align}
%
or equivalently
%
\begin{align}
W(y_{1}, \cdots,-y_{k},\cdots,y_{d})
 = W(y_{1}, \cdots,y_{k},\cdots,y_{d})
   \qquad \textrm{for}\ k=1,\cdots, d.
%
\end{align}
%

%
If the superpotential does not satisfy the above conditions, the extended
supersymmetry is broken.
For instance, let us consider the case that the superpotential $W(y)$
satisfies the relations (\ref{W(-y)=W(y)}) only for the
directions $y_{k_{1}}, \cdots, y_{k_{m}}\ (m < d)$, i.e.
%
\begin{align}
\big(R_{k}W\big)(y) = W(y) \qquad \textrm{for}\ k=k_{1},\cdots,k_{m}\,.
%
\end{align}
%
Then, the supercharges $Q_{k}\ (k \ne k_{1}, \cdots, k_{m})$ and $Q_{d+1}$
in Eq.(\ref{N-extended-even}) become ill-defined and should be removed
from the supersymmetry algebra.
Thus, the supercharges $Q_{k}\ (k = k_{1}, \cdots, k_{m})$ and $Q_{d+2}$
in Eq.(\ref{N-extended-even}) form the $\mathcal{N}=m+1$ extended supersymmetry
for $m=$ odd.
Interestingly, it turns out that the $\mathcal{N}=m+1$ extended supersymmetry
can be enlarged to the $\mathcal{N}=m+2$ extended one for $m=$ even
with an additional supercharge
%
\begin{align}
Q_{(k_{1}\cdots k_{m})}
 \equiv i(-1)^{F} (\one_{2}\otimes i^{m/2}\gamma^{y_{k_{1}}}\cdots\gamma^{y_{k_{m}}})
          R_{k_{1}}\cdots R_{k_{m}}Q\,.
\end{align}
%
The same argument can be also applied to the odd $d$-dimensional case.

%
\section{Summary and discussion}
\label{sec:summary}

In this paper, we have succeeded in constructing the new realization 
of the $\mathcal{N}$-extended quantum-mechanical supersymmetry.
First we considered the hidden $\Ncal=2$
quantum-mechanical supersymmetry  in the KK decomposition of the higher
dimensional Dirac field.
The $\mathcal{N}=2$ supersymmetry shows that the mode function 
$\boldsymbol{f}_{\alpha}^{(n)}$ for the left-handed chiral spinor $\phi^{(n)}_{\alpha}$
and $\boldsymbol{g}_{\alpha}^{(n)}$ for the right-handed chiral spinor 
$\bar{\chi}^{(n)}_{\alpha}$ in Eq.~\eqref{eq:expansion-Dirac} form 
a pair of $\mathcal{N} =2$ supermultiplets.

Surprisingly, we further found that the $\mathcal{N}$-extended quantum-mechanical
supersymmetry with multiple supercharges is hidden in the spectrum.
It turns out that once a single mode function is given, the supercharges
in the $\mathcal{N}$-extended supersymmetry can generate the whole
degenerate mode functions of the KK spectrum on the hyperrectangle
extra dimensions with the boundary conditions~\eqref{eq:BC_g} and~\eqref{eq:BC_f}.

Furthermore we revealed the extended supersymmetry with the superpotential. 
Then, we clarified the conditions for the superpotential 
to preserve or partially preserve the extended supersymmetry. 

%
Our analysis is far from being complete.
We have demonstrated only the case of the boundary conditions given in Eq.~\eqref{eq:BC_g} and~\eqref{eq:BC_f}.
Although the boundary conditions~\eqref{eq:BC_g} and~\eqref{eq:BC_f}
are compatible with the $\mathcal{N}$-extended supersymmetry,
all the supercharges given in Eq.~\eqref{N-extended-even} or~\eqref{N-extended-odd} are not necessarily well-defined for
other types of boundary conditions.
For example, boundary conditions with no reflection symmetry
of $y_{k} \to -y_{k}$, the supercharge $Q_{k}$ will become ill-defined
because $Q_{k}$ includes the reflection operator $R_{k}$ in the
definition of Eq.~\eqref{N-extended-even} or~\eqref{N-extended-odd}.
Thus, it would be of importance to determine how 
the $\mathcal{N}$-extended supersymmetry is broken by boundary conditions.

Allowed boundary conditions on a rectangle (as two dimensional extra dimensions) 
have been classified in Refs.~\cite{Fujimoto2017a,Fujimoto2017b}.
However, general boundary conditions for Dirac fields in arbitrary higher dimensions 
have not been obtained yet, because it is nontrivial to find 
general solutions to Eq.~\eqref{eq:non-local_BC}.
So, it will be interesting to find a class of boundary conditions consistent 
with Eq.~\eqref{eq:non-local_BC}.

Although we have introduced a superpotential into the supercharges, we have not solved
any mass spectrum of concrete models in this paper. 
It would be interesting to try to find a new class of exactly solvable
quantum-mechanical models with the $\mathcal{N}$-extended supersymmetry.
The issues mentioned in this section will be reported in a future work.

\section*{Acknowledgement}

This work is supported
in part by Grants-in-Aid for Scientific Research [No. 15K05055 and No. 25400260 (M.S.)]
from the Ministry of Education, Culture, Sports, Science and Technology (MEXT) in Japan.


\bibliographystyle{elsarticle-num}
\bibliography{Reference_updated,Reference_ver3}

\end{document}